\documentstyle[12pt]{article}

\begin{document}

\author{R. Vilela Mendes \\
Grupo de F\'isica-Matem\'atica, Complexo II - Univ. de Lisboa\\
Av. Gama Pinto, 2 - P1699 Lisboa Codex, Portugal \and Ricardo Coutinho \\
Departamento de Matem\'atica, Instituto Superior T\'ecnico\\
Av. Rovisco Pais, 1096 Lisboa Codex, Portugal}
\title{On the computation of quantum characteristic exponents}
\date{}
\maketitle

\begin{abstract}
A quantum characteristic exponent may be defined, with the same operational
meaning as the classical Lyapunov exponent when the latter is expressed as a
functional of densities. Existence conditions and supporting measure
properties are discussed as well as the problems encountered in the
numerical computation of the quantum exponents. Although an example of true
quantum chaos may be exhibited, the taming effect of quantum mechanics on
chaos is quite apparent in the computation of the quantum exponents.
However, even when the exponents vanish, the functionals used for their
definition may still provide a characterization of distinct complexity
classes for quantum behavior.

Keywords: quantum chaos, characteristic exponents
\end{abstract}

\section{Introduction. Classical and quantum characteristic exponents.}

A notion of {\it quantum characteristic exponent} has been introduced in Ref.%
\cite{Vilela2}, which has the same physical meaning as the corresponding
classical quantity (the Lyapunov exponent). The correspondence is
established by first rewriting the classical Lyapunov exponent as a
functional of densities and then constructing the corresponding quantity in
quantum mechanics. The construction is explained in detail in Ref.\cite
{Vilela1}, where the required functional spaces are identified and the
infinite-dimensional measure theoretic framework is developed. Here we just
recall the main definitions and emphasize some refinements concerning the
support properties of the quantum characteristic exponents, which turn out
to be relevant for the numerical computations of Sect.2.

Expressed as a functional of {\it admissible} $L^1-$densities, the classical
Lyapunov exponent is\cite{Vilela1} 
\begin{equation}
\label{1.1}\lambda _v=\lim _{n\rightarrow \infty }\frac 1n\log \left|
-v^i\frac \partial {\partial x^i}D_{\delta _x}\left( \int d\mu (y)\smallskip%
\ y\smallskip\ P^n\rho (y)\right) \right| 
\end{equation}
where $\rho $ is an initial condition density, $P$ the Perron-Frobenius
operator, $x$ a generic phase-space coordinate, $v$ a vector in the tangent
space, $\mu $ the invariant measure and $D_{\delta _x}$ the Gateaux
derivative along the generalized function $\delta _x$. The possibility to
define Gateaux derivatives along generalized functions with point support
and the need for a well-defined $\sigma $-additive measure in an
infinite-dimensional functional space lead almost uniquely to the choice of
the appropriate mathematical framework, that is, {\it admissible densities}
are required to belong to a nuclear space. Being ergodic invariants, the
Lyapunov exponents exist on the support of a measure. In the nuclear space
framework, measures with support on generalized functions, which are in
one-to-one correspondence with the usual measures in phase space, may be
constructed by the Bochner-Minlos theorem\cite{Vilela1}.

To construct, in quantum mechanics, a quantity with the same operational
meaning as (\ref{1.1}) let $U^n$ ($n$ continuous or discrete) be the unitary
operator of quantum evolution acting on the Hilbert space $H$ and $%
\widetilde{X}$ a self-adjoint operator in $H$ belonging to a commuting set $S
$. For definiteness $\widetilde{X}$ is assumed to have a continuous
spectrum, to be for example a coordinate operator in configuration space.
One considers, as in the classical case, the propagation of a perturbation $%
\partial _i\delta _x$ , where by $x$ we mean now a point in the spectrum of $%
\widetilde{X}$. 
\begin{equation}
\label{1.2}v^iD_{\partial _i\delta _x}\left( U^n\Psi ,\widetilde{X}U^n\Psi
\right) =2Re\bigskip\ v^i\frac \partial {\partial x^i}<\delta _x,U^{-n}%
\widetilde{X}U^n\Psi > 
\end{equation}
For the proper definition of the right-hand side of (\ref{1.2}) one requires 
$\Psi \in E$ to be in a Gelfand triplet\cite{Gelfand}%
$$
E^{*}\supset H\supset E 
$$
By $<\delta _x|$ or $<x|$ we denote a generalized eigenvector of $\widetilde{%
X}$ in $E^{*}$. Notice also that $U^n$, being an element of the
infinite-dimensional unitary group, has a natural action both in $E$ and $%
E^{*}$\cite{Hida}. One obtains then the following definition of {\it quantum
characteristic exponent} 
\begin{equation}
\label{1.3}\lambda _{v,x}=\lim \sup _{n\rightarrow \infty }\frac 1n\log
\left| \textnormal{Re}\bigskip\ v^i\frac \partial {\partial x^i}<\delta _x,U^{-n}%
\widetilde{X}U^n\Psi >\right| 
\end{equation}

The support properties of this quantum version of the Lyapunov exponent have
to be carefully analyzed. In Eq.(\ref{1.3}), $\Psi $ defines the state
which, in quantum mechanics, plays the role of a (non-commutative) measure%
\cite{Connes1}. The quantum exponent may depend on the state, but the
measure that, as in the classical case, provides its support is not the
state but a measure in the space of the perturbations of the initial
conditions, that is, in the space where the Gateaux derivative operates. In
the classical case these two measures coincide, in the sense that to which
invariant measure in phase-space corresponds an infinite-dimensional measure
in the space of generalized functions\cite{Vilela1}. In the quantum case,
however, the two entities are different, the second one being the measure on
the spectrum of $\widetilde{X}$ induced by the projection-valued spectral
measure and the state, that is

\begin{equation}
\label{1.4}\nu (\Delta x)=(\Psi ,\int\limits_{\Delta x}dP_x\Psi ) 
\end{equation}

A particular case where an infinite-dimensional measure-theoretical setting,
similar to the classical one, may be used to define the quantum exponents%
\cite{Vilela1}, is when the quantum evolution is implemented by substitution
operators in configuration space, as in some sectors of the configurational
quantum cat\cite{Weigert1}\cite{Vilela2}\cite{Weigert2}. However this
formulation is not very useful in general and the {\it state plus
spectrum-measure} framework seems to be the one that has general validity.
In this framework the following existence theorem holds

\underline{{\em Theorem}}: Let $\widetilde{X}$ be a self-adjoint operator, $E
$ a test function space in a Gelfand triplet containing the generalized
eigenvectors of $\widetilde{X}$ in its dual $E^{*}$ and $\Psi \in E$. Then
if $U^n\Phi \in E$ and $\widetilde{X}\Phi \in E$ ($\forall \Phi \in E$, $%
\forall n\in Z$) and the following integrability condition is fulfilled 
\begin{equation}
\label{1.4a}\left| \int d\nu (x)\log \frac{\left| \textnormal{Re}v^i\partial
_{x_i}<x|U^{-1}\Phi >\right| }{\left| \textnormal{Re}v^i\partial _{x_i}<x|\Phi
>\right| }\right| <M 
\end{equation}
$\forall \Phi \in E$, the limit in Eq.(\ref{1.3}) exists as a $L^1(\nu )-$%
function, that is , the average quantum characteristic exponent is defined
for any measurable set in the support of $\nu $.\bigskip\ 

\bigskip 

Proof:

We write Eq.(\ref{1.3}) as 
\begin{equation}
\label{1.4b}\lambda _v(x)=\lim \sup _{n\rightarrow \infty }\frac
1n\sum_{n=0}^{n-1}\log \frac{\left| \textnormal{Re}v^i\partial _{x_i}<x|U^{-n+k}%
\widetilde{X}U^{n-k}\Phi >\right| }{\left| \textnormal{Re}v^i\partial
_{x_i}<x|U^{-n+k+1}\widetilde{X}U^{n-k-1}\Phi >\right| } 
\end{equation}
Then from the integrability condition (\ref{1.4a}) the integral of the
sequence in the right-hand side of Eq.(\ref{1.4b}) is bounded and the
Bolzano-Weierstrass theorem insures the existence of the $\lim \sup $.
Therefore $\lambda _v(x)$ is well defined as a $L^1(\nu )-$function. $\Box $

Notice that we really need the $\lim \sup $ in the definition of the
characteristic exponent because we have no natural $U$-invariant measure in $%
E$ to be able, for example, to apply Birkhoff or Kingman's theorem and prove 
$\lim \sup $=$\lim \inf $. Also the sense in which the measure $\nu $
provides the support for the quantum characteristic exponent is different
from the classical ergodic theorems. We have not proven pointwise existence
of the exponent a. e. in the support a measure. What we have obtained here
is the possibility to define an average quantum characteristic exponent for
arbitrarily small $\nu -$measurable sets.

Other definitions of characteristic exponents in infinite-dimensional spaces
have been proposed by several authors\cite{Ruelle2} \cite{Vilela3} \cite
{Haake} \cite{Zycz} \cite{Majewski} \cite{Emch}. They characterize several
aspects of the dynamics of linear and non-linear systems. The definition
discussed here, proposed for the first time in \cite{Vilela2}, seems however
to be the one that is as close as possible to the spirit of the classical
definition of Lyapunov exponent.

Like the classical Lyapunov exponent the quantum analogue (\ref{1.3}) cannot
in general be obtained analytically. There is however a non-trivial example
where it can. This is the configurational quantum cat introduced by Weigert%
\cite{Weigert1}\cite{Weigert2}. The phase space of this model is $T^2\times
R^2$. A mapping similar to the classical cat operates as a quantum kick in
the configuration space $T^2$, and the rest of the Floquet operator is a
free evolution. This system has the appealing features of actually
corresponding to the physical motion of a charged particle on a torus acted
by an impulsive electromagnetic field and, as show by Weigert\cite{Weigert2}%
, to be exactly solvable.

The Floquet operator is 
\begin{equation}
\label{1.5}U=U_FU_K 
\end{equation}
\begin{equation}
\label{1.6}U_F=\exp [-i\frac T2\widetilde{p}^2];\bigskip\ U_K=\exp [-\frac
i2(\widetilde{x}\cdot V\cdot \widetilde{p}+\widetilde{p}\cdot V\cdot 
\widetilde{x})] 
\end{equation}
$U_F$ is a free evolution and $U_K$ a kick that operates in a simple manner
on momentum eigenstates and on (generalized) position eigenstates 
\begin{equation}
\label{1.7}U_K\left| p\right\rangle =\left| M^{-1}p\right\rangle 
\end{equation}
\begin{equation}
\label{1.8}U_K\left| x\right\rangle =\left| M\textnormal{ }x\right\rangle 
\end{equation}
$M$ being an hyperbolic matrix with integer entries and determinant equal to
1. The momentum has discrete spectrum, $p\in (2\pi Z)^2$.

To compute the quantum characteristic exponent (Eq.(\ref{1.3})), let the
operator $\widetilde{X}$ be 
\begin{equation}
\label{1.9}\widetilde{X}=\sin (2\pi l\cdot x) 
\end{equation}
$l\in Z^2$. This operator has the same set of generalized eigenvectors as
the position operator $\widetilde{x}$. To construct the measure $\nu $ (Eq.(%
\ref{1.4})) in the spectrum of the operator $\widetilde{X}$ we cannot use
the energy eigenstates $\mid P\alpha \rangle $ because they are not
normalized. However all one requires is invariance of the measure, and using
the (normalizable) momentum eigenstates one such measure is obtained. 
\begin{equation}
\label{1.10}\nu (A)=\langle p\mid \int_Adx\mid x\rangle \langle x\mid
p\rangle 
\end{equation}
This invariant measure in this case happens to be simply the Lebesgue
measure in $T^2$.

Defining 
\begin{equation}
\label{1.11}\gamma _n(x)=\langle x\mid U^{-n}\widetilde{X}U^n\mid p\rangle 
\end{equation}
the result for the quantum characteristic exponent is 
\begin{equation}
\label{1.12}
\begin{array}{c}
\lambda _v=\lim \sup _{n\rightarrow \infty }\frac 1n\log ^{+}\left|
Rev^i\frac \partial {\partial x^i}\gamma _n(x)\right| \\ 
=\lim \sup _{n\rightarrow \infty }\frac 1n\log ^{+}\left| v^i(2\pi
M^nl)_i\{\cos \theta _n(p,l,x)+\cos \theta _n(p,-l,x)\}\right| 
\end{array}
\end{equation}
with 
$$
\theta _n(p,l,x)=\frac T2\left(
\sum_{k=0}^{n-1}(M^{-k}p)^2+\sum_{k=0}^{n-1}(M^k(2\pi l+M^{-n}p))^2+x\cdot
(2\pi M^nl+p)\right) 
$$
For the $\lim \sup $ the cosine term plays no role and finally 
\begin{equation}
\label{1.13}\lambda _v=\lim \sup _{n\rightarrow \infty }\frac 1n\log \left|
v^i(M^nl)_i\right| 
\end{equation}
The characteristic exponent is then determined from the eigenvalues of the
hyperbolic matrix $M$ and is the same everywhere in the support of the
measure $\nu $. If $\mu _1$, $\mu _2$ ($\mu _1>\mu _2$) are
the eigenvalues of $M$, one obtains $\lambda _v=\log \mu _1$ for a generic
vector $v$ and $\lambda _v=\log \mu _2$ iff $\nu $ is orthogonal to the
eigenvector associated to $\mu _1$. Hence, in this case, one obtains a
positive quantum characteristic exponent whenever the corresponding
classical Lyapunov exponent is also positive.

This exact example will be used in Sect.2 as a testing ground for the
numerical algorithm and an illustration of the kind of precision problems
and support properties to be expected when computing quantum characteristic
exponents.

In the numerical calculation of the quantum characteristic exponents two
delicate points are identified. The first is that the calculation requires a
high degree of precision, because, if the exponent is positive, the
derivative of $U^{-n}\widetilde{X}U^n\Psi (x)$ grows very rapidly with $n$.
Therefore in the positive exponent case acceptable statistics is only
obtained by taking average values over the configuration space. Second, if
the situation is as in the classical case where different invariant measures
coexist in phase space, the quantum exponent may depend on the state $\Psi $
used to define the measure $\nu $ in the spectrum of $\widetilde{X}$.
Therefore, in all rigor, one should first construct stationary states and
then study the $\Psi -$dependence of the quantum exponent. Such study has
not yet been carried out and, in the calculations below, a flat wave
function is used as the initial state.

\section{Numerical computation of quantum characteristic exponents}

For kicked quantum systems corresponding to the Hamiltonian 
\begin{equation}
\label{2.1}H=H_0+V(x)\sum_j\delta (t-j\tau ) 
\end{equation}
the Floquet operator is 
\begin{equation}
\label{2.2}U=e^{-iV(x)}e^{-i\tau H_0(\frac \partial {\partial x_i})} 
\end{equation}
in units where $\hbar =1$. For the computation of the action of $U$ on a
wave function $\psi (x)$, a fast Fourier transform algorithm $F$ and its
inverse $F^{-1}$ are used 
\begin{equation}
\label{2.3}U\psi (x)=e^{-iV(x)}F^{-1}e^{-i\tau H_0(ik)}F\psi (x) 
\end{equation}
In this way one obtains a uniform algorithm for any potential. In the
configurational quantum cat, Eq.(\ref{1.6}), the computation is similar with
the multiplicative kick $e^{-iV(x)}$ replaced by the substitution operator 
\begin{equation}
\label{2.4}\psi (x)\rightarrow \psi (M^{-1}x) 
\end{equation}
The quantum characteristic exponent is obtained from the calculation of 
\begin{equation}
\label{2.5}\partial _xU^{-n}\widetilde{X}U^n\psi (x) 
\end{equation}
in the limit of large $n$. The precision of the algorithm is checked by
insuring that 
\begin{equation}
\label{2.6}\left| \left( U^{-n}U^n-1\right) \psi (x)\right| <\epsilon 
\end{equation}
for a small $\epsilon $, and that the finite difference used to compute the
derivative in (\ref{2.5}) does not approach the maximum possible value
allowed by the discretization.

\subsection{The configurational quantum cat}

Here the configuration space is the 2-torus $T^2$, the Floquet operator is
the one of Eq.(\ref{1.6}), and the kick is a substitution operator with
matrix 
\begin{equation}
\label{2.7}M=\left( 
\begin{array}{cc}
1 & 1 \\ 
1 & 2
\end{array}
\right) 
\end{equation}
Numerically we have computed the quantities 
\begin{equation}
\label{2.8}\frac 1n\left\langle D_n-D_0\right\rangle =\frac 1n\left\langle
\log \frac{\left| Re\bigskip\ v^i\frac \partial {\partial x^i}\left( U^{-n}%
\widetilde{X}U^n\Psi \right) (x)\right| }{\left| Re\bigskip\ v^i\frac
\partial {\partial x^i}\left( \widetilde{X}\Psi \right) (x)\right| }%
\right\rangle _{T^2}
\end{equation}
$\widetilde{X}$ being the operator in (\ref{1.9}). The initial wave function
is $\Psi (x)=1$ and the average is taken over the whole of configuration
space. It turns out that, in this case, the derivative in the numerator of (%
\ref{2.8}) grows so fast at some points that one reaches, after a few
iterations, the maximum difference (2 in this case) for the wave function at
two nearby points in the discretization grid. When this happens the
calculation cannot be reliably taken to higher $n$ with that discretization.
In the numerical calculation of the classical Lyapunov the computation
becomes a local evaluation at each step by rescaling the transported tangent
vector. Here, because of the linearity of matrix elements and quantum
evolution, a similar procedure is not possible and one has to carefully
control the growth of the quantities in (\ref{2.8}). To improve statistics
the average over configuration space has been taken. This can be safely done
in this case because we know exactly the supporting measure (\ref{1.10}),
but in general there will be no guarantee that the supporting spectral
measure is uniform. In any case average quantities like (\ref{2.8}) are
exactly we expect to be able to compute reliably.

Fig.1 shows the evolution of $\frac 1n\left\langle D_n-D_0\right\rangle $
obtained with a discretization grid of 292681 points in the unit square, for
two different directions $\nu $. The calculation was interrupted when the
local finite differences reached one half of the maximum. The lines are fits
to the points constrained to approach the same value at large $n$. The
resulting numerical estimate for the largest quantum characteristic exponent
is 0.95. The exact value obtained from (\ref{1.13}) is 0.9624.

\subsection{Quantum kicked rotators}

The configuration space is the circle $S^1$, 
\begin{equation}
\label{2.9}V(x)=q\cos (2\pi x) 
\end{equation}
$x\in [0,1)$ and for $H_0$ the following two possibilities were explored 
\begin{equation}
\label{2.10}
\begin{array}{c}
H_0^{(1)}=-\frac 1{2\pi } 
\frac{d^2}{dx^2} \\ H_0^{(2)}=-2\pi \cos \left( \frac 1{2\pi i}\frac
d{dx}\right) 
\end{array}
\end{equation}
The operator $\widetilde{X}$ is 
\begin{equation}
\label{2.11}\widetilde{X}=\sin (2\pi x) 
\end{equation}
The quantity that is numerically computed is

\begin{equation}
\label{2.12}\left\langle D_n\right\rangle =\left\langle \log \left| Re%
\bigskip\ \frac \partial {\partial x}\left( U^{-n}\widetilde{X}U^n\Psi
\right) (x)\right| \right\rangle 
\end{equation}
and, in all cases, one starts from a flat initial wave function. In both
cases and for the very many values of $q$ that were studied, this quantity
seems either to stabilize or to have a very small rate of growth for large $n
$. Fig.2, for example, shows the results for the $H_0^{(1)}$ case with $\tau
=\frac{\sqrt{5}}2$ and $q=5$. The (numerical) conclusion is that the quantum
characteristic exponent vanishes. This conclusion does not seem to be a
numerical artifact because the discretization grid for the fast Fourier
transform has always been chosen sufficiently small to insure a small local
finite difference for all iterations. For example in the example shown in
Fig.2, the grid has 4096 points which keeps observed local differences below
0.1. Also the vanishing of the quantum characteristic exponent in quantum
kicked rotators is not dependent on the phenomena of localization because
also for quantum resonances it may exactly be shown to vanish\cite{Vilela2}.

In Fig.2 $\left\langle D_n\right\rangle $ seems to tend to a constant at
large $n$. In other cases very slow rates of growth are observed. This is
shown in Figs.3a,b for the $H_0^{(2)}$ case with $\tau =\frac{\sqrt{5}}2$
and $q=11$.

\section{Conclusions}

Both the classical Lyapunov exponent (\ref{1.1}) and its quantum counterpart
(\ref{1.3}), measure the exponential rate of separation of matrix elements
of $\widetilde{X}$ when the density (or the wave function) suffers a $\delta
_x^{^{\prime }}$ perturbation, $x$ being a point in the spectrum of $%
\widetilde{X}$. The configurational quantum cat example shows that there are
instances of {\it true quantum chaos}, in the sense of exponential growth of
the matrix element separation. However, as the numerical study of the
quantum kicked rotators seems to show, exponential growth may be rather
exceptional in quantum mechanics. Furthermore the taming effect of quantum
mechanics on exponential chaos goes deeper than the phenomenon of
localization, because also for quantum resonances, where no localization is
present, the quantum characteristic exponent vanishes.

Although distinct from one another, all known ways that now exist to
approach the problem of quantum chaos, seem to agree in one point, namely
that quantum mechanics has a definite taming effect on chaos. This is now
probably the main issue in quantum chaos, not only from the theoretical
point of view, but also in the context of quantum control. Even if quantum
characteristic exponent, as defined in (\ref{1.3}) might be zero in most
quantum systems, the rate of growth index $D_n$ or its average $\left\langle
D_n\right\rangle $ might still be useful as a characterization of quantum
dynamics because, even if weaker than exponential, a growth of this quantity
would still be an indication of sensitivity to initial conditions. In
particular, as suggested by the numerical results, subexponential rates of
growth might characterize distinct complexity classes of quantum evolution.

\section{Figure captions}

Fig.1 - Calculation of $\frac 1n\left\langle D_n-D_0\right\rangle $, Eq.(\ref
{2.8}), in the configurational quantum cat for two orthogonal directions $%
\nu $ and a fit constrained to the same limit at large $n$.

Fig.2 - $\left\langle D_n\right\rangle $, Eq.(\ref{2.12}), for the quantum
standard map at $q=5$, $\tau =\frac{\sqrt{5}}2$.

Fig.3 - (a) $\left\langle D_n\right\rangle $, Eq.(\ref{2.12}), for a kicked
rotator with kinetic Hamiltonian $H_0^{(2)}$ at $q=11$, $\tau =\frac{\sqrt{5}%
}2$ ; (b) the same scaled by $\log (\log (n+1))$.


\begin{thebibliography}{99}
\bibitem{Vilela2}  R. Vilela Mendes; Phys. Lett. A171, 253 (1992).

\bibitem{Vilela1}  R. Vilela Mendes; in {\sl Chaos - The Interplay between
Stochastic and Deterministic Behaviour}, page 273, P. Garbaczewski and A.
Weron (Eds.), Springer Lecture Notes in Physics no. 457, Springer, Berlin
1995.

\bibitem{Gelfand}  I. M. Gelfand and N. Ya. Vilenkin; {\sl Generalized
functions}, vol. 4, Academic Press, New York 1964.

\bibitem{Hida}  T. Hida; {\sl Brownian Motion}, Springer, Berlin 1980.

\bibitem{Connes1}  A. Connes; {\sl Noncommutative geometry}, Academic Press
1994.

\bibitem{Weigert1}  S. Weigert; Z. Phys. B - Condensed Matter 80, 3 (1990).

\bibitem{Weigert2}  S. Weigert; Phys. Rev. A48, 1780 (1993).

\bibitem{Ruelle2}  D. Ruelle; Ann. Math. 115, 243 (1982).

\bibitem{Vilela3}  R. Vilela Mendes; J. Phys. A: Math. Gen. 24, 4349 (1991).

\bibitem{Haake}  F. Haake, H. Wiedemann and K. Zyczkowski; Ann. Physik 1,
531 (1992).

\bibitem{Zycz}  K. Zyczkowski, H. Wiedeman and W. Slomczynski; Vistas in
Astronomy 37, 153 (1993).

\bibitem{Majewski}  W. Majewski and M. Kuna, J. Math. Phys. 34, 5007 (1993).

\bibitem{Emch}  G. G. Emch, H. Narnhofer, W. Thirring and G. L. Sewell; J.
Math. Phys. 35, 5582 (1994).
\end{thebibliography}
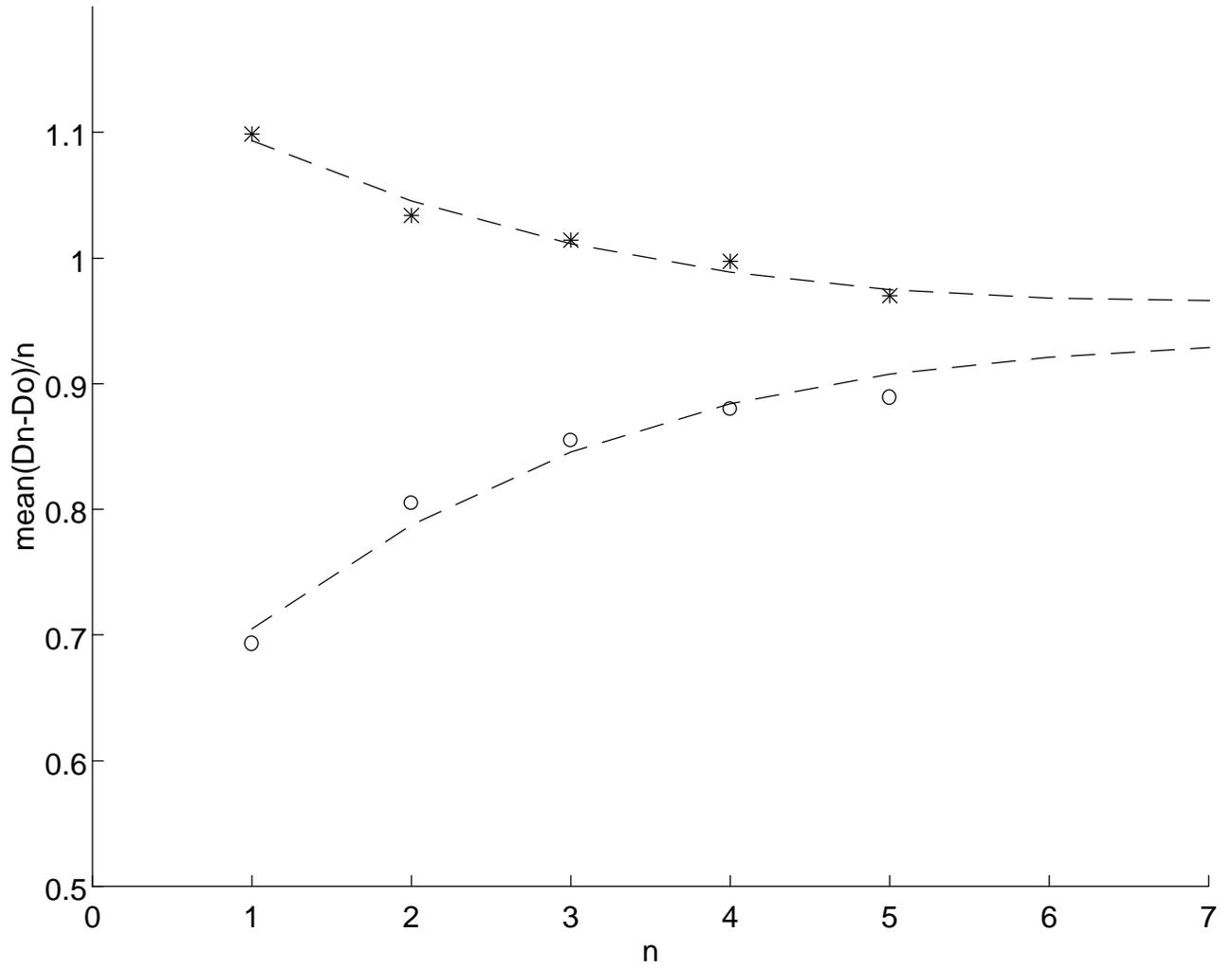
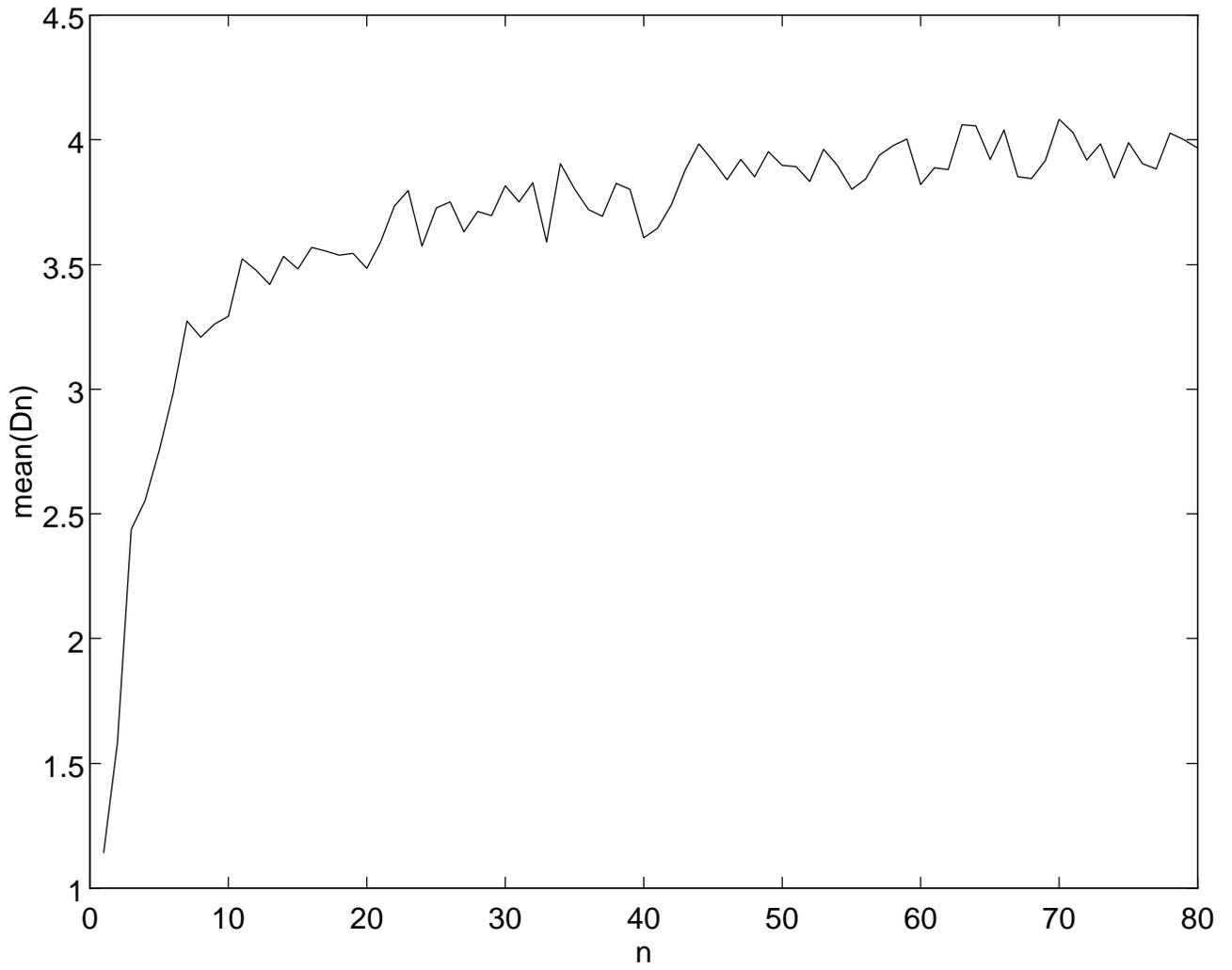
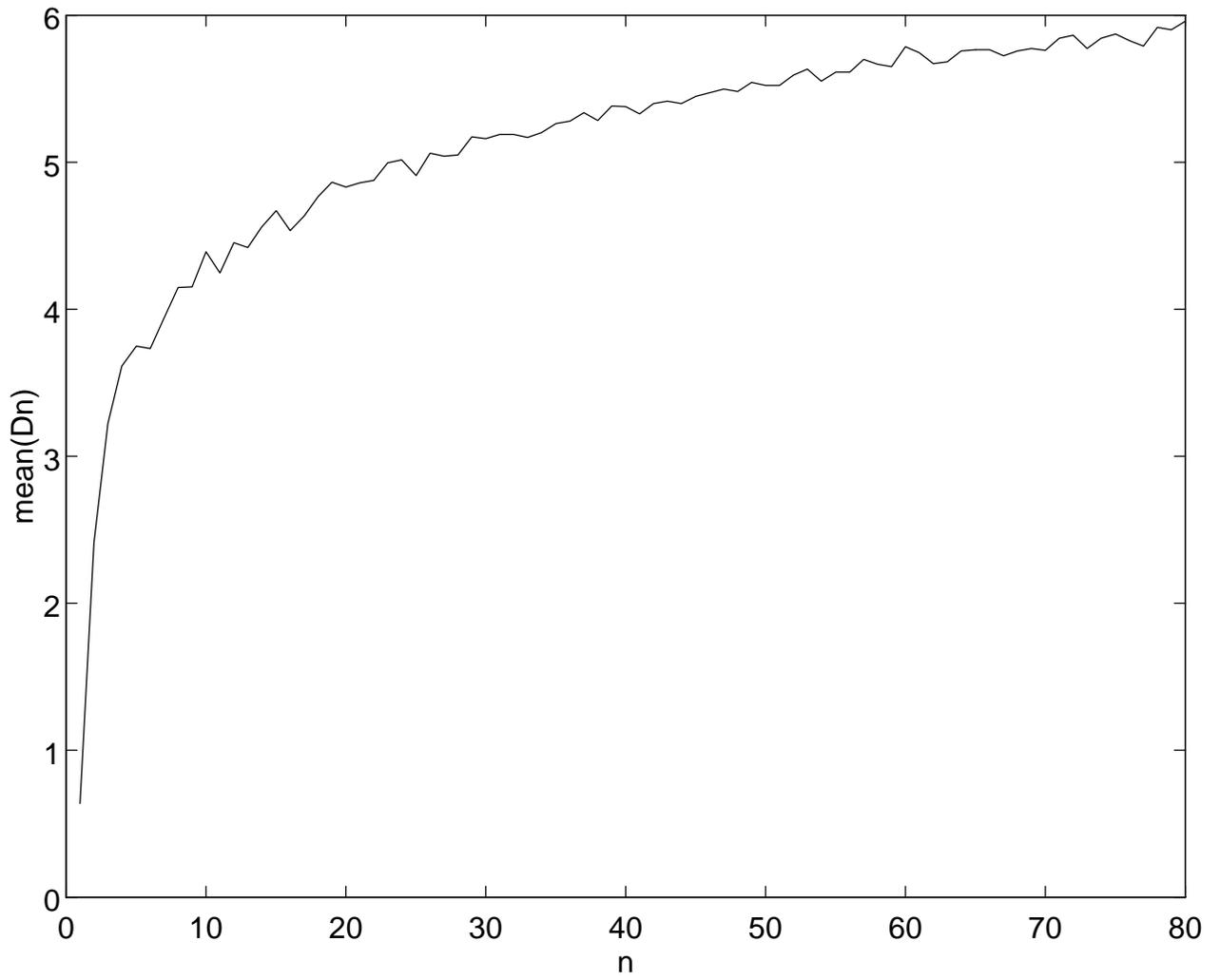
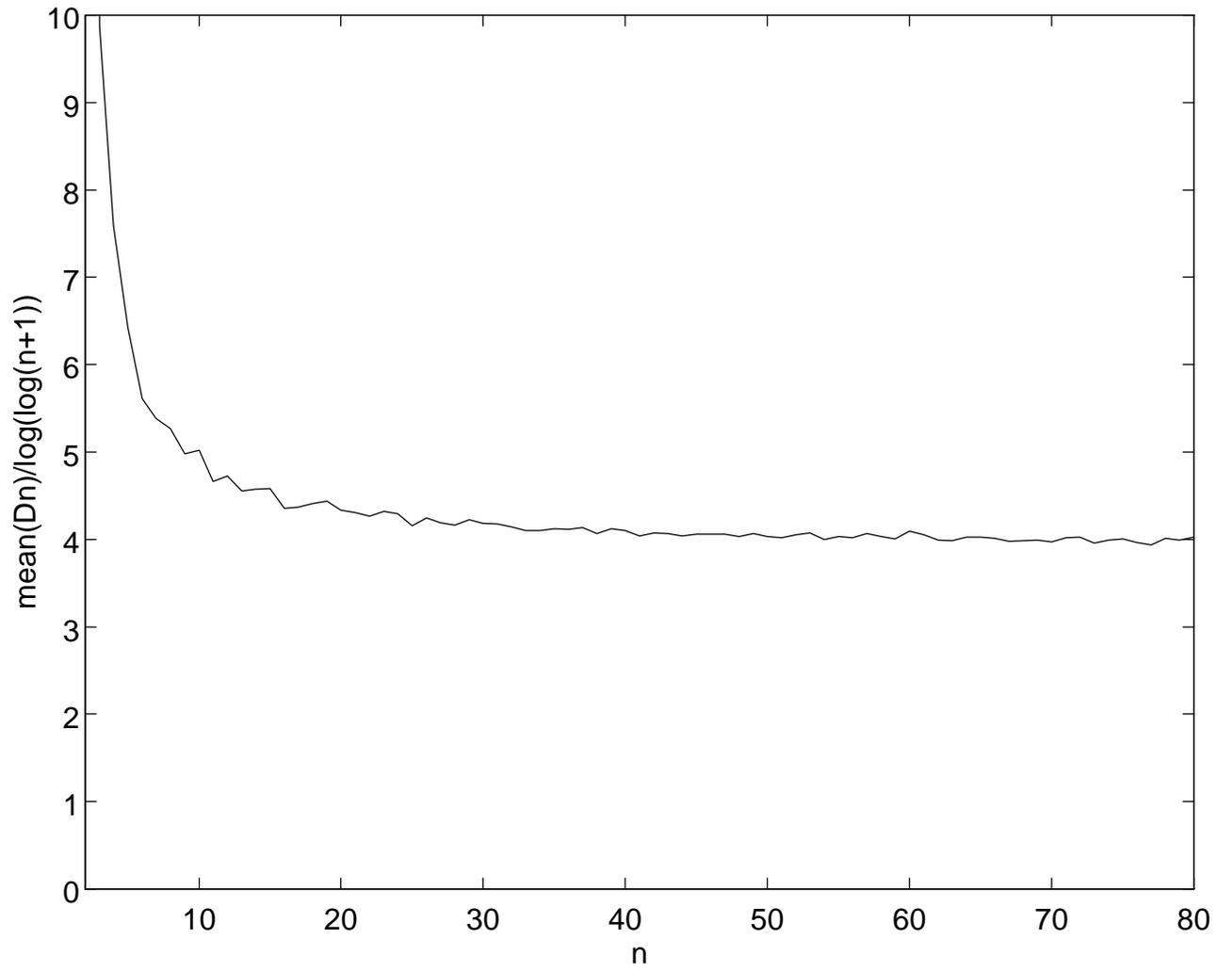
\end{document}